# AN INNOVATIVE ECO-SYSTEM FOR ACCELERATOR SCIENCE AND TECHNOLOGY

C. Darve[†], J. Andersen, S. Salman, European Spallation Source, ERIC, Lund, Sweden
M. Stankovski, Lund Institute of Advanced Neutron and X-ray Science, Lund, Sweden
B. Nicquevert[1], S. Petit, CERN, Geneva, Switzerland
[1]also at European Spallation Source, ERIC, Lund, Sweden

## Abstract

The emergence of new technologies and innovative communication tools permits us to transcend societal challenges. While particle accelerators are essential instruments to improve our quality of life through science and technology, an adequate ecosystem is essential to activate and maximize this potential. Research Infrastructure (RI) and industries supported by enlightened organizations and education, can generate a sustainable environment to serve this purpose. In this paper, we will discuss state-of-the-art infrastructures taking the lead to reach this impact, thus contributing to economic and social transformation.

## AN EMERGING ECOSYSTEM

Particle accelerators and Large Scale Research Infrastructures (LSRIs) - which are crucial for providing impactful science for society - have for nearly nine decades experienced continued growth in their energy reach. The need to develop excellent science has thus driven continuous innovation and evolution in acceleration techniques. Particle accelerators can cover a broad range of purposes, from medical applications, radiotherapy, defence, hard to soft matter research, to large-scale research infrastructure for high-energy physics or basic science. Industries benefit from particle accelerator fabrication, operation and breakthroughs resulting from large neutron and light sources with the typical average proportion of industry use of neutron and light-based LSRIs in Europe today being 20%-40% in collaboration with academia, and 1%-10% purely proprietary use, depending on the type of LSRI [1-5].

An overwhelming proportion of the private sector engagement is done in collaboration with others. We will in this contribution use the emerging ecosystem around the European Spallation Source ERIC (ESS) and its context in comparison to other LSRIs as an example.

Technology and knowledge transfer are the pillars of the development of this innovative ecosystem, using scientific communication, education and collaboration as their vectors. University programs and advanced studies institutes can train human resources to build, operate and get the most use out of scientific infrastructures, wherein the context of ESS and MAX IV Laboratory in Sweden, the Lund Institute for advanced Neutron and X-ray Science (LINXS) is one example that fills this niche for academic and industrial use. The similarly named danish Linking Industry to Neutrons and X-rays (LINX) association fills the niche of promoting industrial use specifically. Figure 1 below shows a constellation of research facilities, academia and research institutes working in any given ecosystem and schematically how Research laboratories, Industry and Universities can shape an ecosystem, its vectors, and the functions typically performed by them.

In the framework of research laboratories like CERN, where the world wide web was born 30 years ago, or recent neutron and light sources, like ESS and MAX IV, Research Infrastructures and academic environments evolve to support international "Big Science" projects like the LHC or ESS instruments. (For a detailed overview at the EU level we refer to the ESFRI roadmap [6-7].) These world-leading institutes support scientific and technological disciplines, which enable ground-breaking research and develop our collective capability to answer societal challenges [8-9].

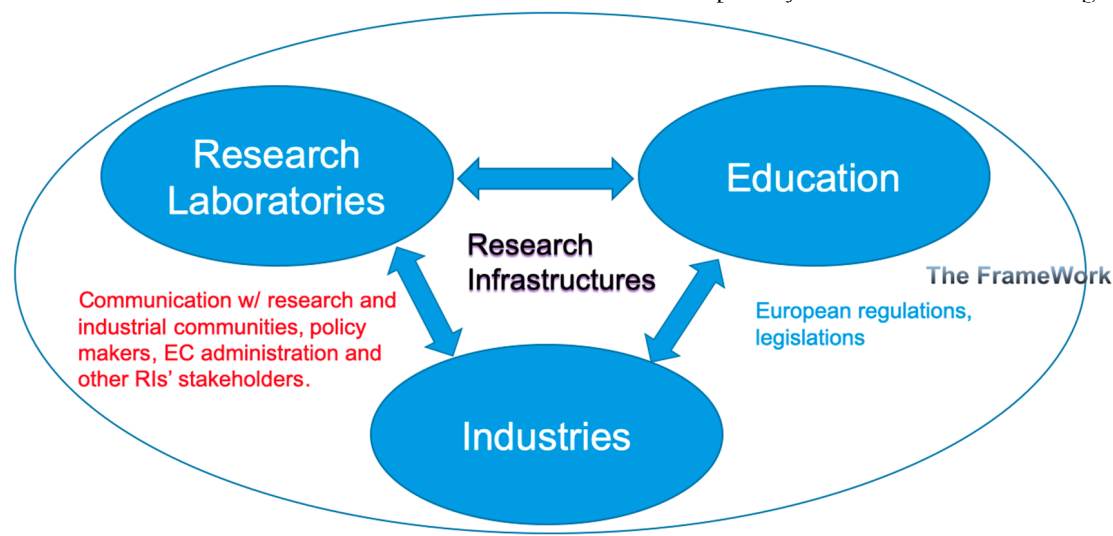

[†] christine.darve@ess.eu

Figure 1: Typical ecosystem map for a Research Infrastructure

Figure 2 shows an example of the synergies between 4 main stakeholder groups, that together empower solution driven and results focused execution of projects. Innovative ecosystems like these, enable science for society. Existing ecosystems, such as LINXS (swedish) [10], LEAPS [11], CERIC [12], EIROforum [13], LENS [14] or LINX (danish) [15], revive forum and communication with research and industrial communities, policymakers, EC administration and other RIs' stakeholders. Venture capital funds beyond traditional project funding are crucial to realize and execute the commercial value and impact.

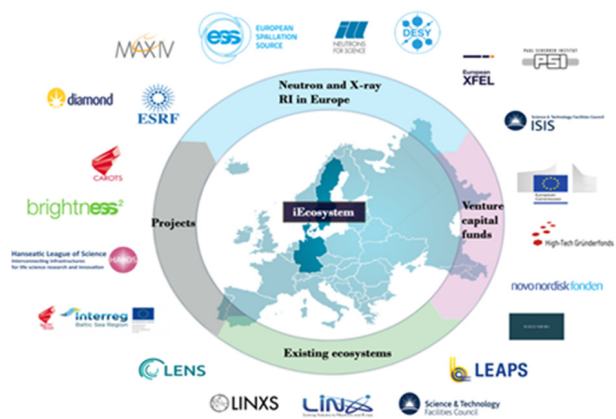

Figure 2: Specific innovative ecosystem

Due to the large scale of the supply chain required to construct and operate those giant RI's, adapted collaborative models are essential. The In-Kind model permits to engage financially the skilled personnel and procure performant equipment.

## THE DATA PROCESS

The recent evolution of technologies and Information and communications technology (ICT) opens the doors to innovative tools development. Since the beginning of the pandemic, the digital online world has matured significantly, embracing technical opportunities and performance. Access to knowledge has become simpler and large quantities of data can be exchanged in a timely manner. Smart/agile data storage, formatting and standardization are required to establish a FAIR data principle [16].

From the web to distributed mega clouds, and from advanced machine learning and artificial intelligence, from advanced handheld computer power to quantum computing, these are all building blocks of the vast data processing and distribution opportunities around the corner [17-18]. It is how all these capabilities and technologies are connecting and complementing each other, that the true value and impact will arise.

Whether the evolution takes a semantic or cognitive route or a combination of both, the future possibilities are probably beyond our imagination.

Broad types of data can be processed to construct, operate and maintain large RI. Processes will ensure that the information collected can be wisely organized (with insight) and permit a worthwhile/behove impact.

## ESS INNOVATION DEVELOPMENT

Research Infrastructures continue to innovate and find solutions for their special needs. The construction of the ESS provides many innovation and technology transfer opportunities and innovation takes place in each technical areas:

- The Accelerator will be the world's most powerful proton accelerator, and as such, a technical challenge to design.
- The Target concept will secure the best scientific performance combined with safety and minimum environmental impact.
- The 15 neutron instruments are versatile, and most serve several different scientific communities.
- The Integrated Control System (ICS) for the European Spallation Source is a complex network of hardware, software and configuration databases.
- The Data is the generic currency of the complex system control and the scientific results
  o Scientific Data where the volume, quality, storage and dissemination is challenging.
  o Control Data from sensors and systems throughout the whole technical installation.

Business cases describe the innovative solutions developed for and at ESS, their application, operating principles, and potential marketability. ESS catalogue of innovative technologies, enabling commercialization is growing, e.g. non-invasive beam diagnostics equipment invention [19-22].

## INNOVATIVE TOOLS APPLICABILITY

Data processing and the application of new tools are important drivers of innovative ecosystems. The following examples, list existing tools expanding RI performance.

### Bridging Construction and Technology

**ESS Management System** The green-field ESS is based on an Integrated Management System set of controlled documents used by an organization to ensure that it can fulfil the tasks required to achieve its objectives. It is a single system designed to manage multiple domains of an organisation, such as quality, environmental and health and safety management in compliance with EU legislation and regulations.

**Track It** A tool bringing several communities of engineers together in order to improve the way technical stops are handled [23]. Acting as an aggregator, it provides a hub for information and people, helping tracking modifications of machines, such as the LHC. It covers various phases of change: preparation, implementation on the field and follow-up, as seen from various angles, like configuration management, machine integration, coordination and planning, safety. It is also helpful for making sure that nothing is forgotten between two technical stops, for instance when an activity needed to be postponed. Its strong integration with pre-existing systems, like the CERN engineering documentation management system (EDMS) ensures that crucial information is not duplicated however made available at the right place and moment [24]. Track It can also be seen as a tool made for capturing knowledge, as it hosts information describing why some decisions were made.

**Remote access operation** Nowadays, secured tools permit researchers to operate RIs remotely, running crucial experiments or COVID-related experiments from home during the pandemic. MAXIV Laboratory and Diamond LS have allowed up to 50% remote operation.

**Virtual room** The green-field ESS has used state-of-the-art engineering way-of-working. Virtual Reality (VR) is used for the installation of equipment is following the Spatial Integration according to the 3D Master Model of the ESS plant layout.

**Augmented reality** CERN integrated safety systems use VR and Augmented Reality (AR) during planned and emergency maintenance in extreme environments. Cameras mounted on the helmet permit to monitor the working area and complete maintenance in controlled areas [25].

**Innovation applied to training** One of the great challenges in Learning & Training is the question of which training is suitable for which person and how this knowledge can be made easily accessible to the user (learner). To address this challenge, ESS Safety Training has developed a "Role-Based Approach to Training". This approach has been conceptually completed and is currently in primary application with some selected ESS stakeholders.

*Educating and Engaging*

The goal of the following initiatives is to catalyze the development of world-class institutions through the production of high–quality scientists and engineers to stimulate economic growth, employment creation, entrepreneurial and leadership capacities to solve problems thus contributing to economic and social transformation, whilst transmitting the thrill and curiosity of these disciplines.

**Massive Open Online Courses (MOOC)** Advanced and innovative MOOCs have been developed in the framework of NPAP and ARIES, financially supported by European grants [26-27], teaching today accelerator physics and technologies to currently more than 10,000 learners.

**On-line lectures** Since the beginning of the pandemic Online communication, has been integrated into our way-of-living. Those on-line courses are sustainable and open-source training the future scientists anywhere in the world. For instance, the ASP has replaced the bi-annual summer school event with 60 Online lectures, including a series dedicated to light sources and neutron sources [28-29].

**Public engagement** Similar tools are developed to reach the younger generations and a broader audience. For instance, an initiative to increase teachers' confidence and subject knowledge in key areas of the curriculum has been addressed by ESS in collaboration with other European Research Laboratories. Concrete examples of various particle accelerators provide relevant context, along with curriculum-linked associated classroom activities.

Knowledge transfer and public awareness are the vectors of a successful education, they can contribute to communities development, reaching the realms of humanity, e.g. life colloquiums in and for developing scientific communities permit to foster knowledge [30-31]

## THE NEXT STEPS

While high-profile impactful business models are one of the ultimate goals of Big Science and accelerator-based RI's, the implementation of specific high-quality scientific tools is essential. Hence, we propose the following step to gain some insight into the innovative ecosystem. The next steps considered are as follows:

- Complete market survey of existing innovative tools, constraints and capacity from existing laboratories.
- Build synergies and exchange good practices
- Develop a proof of concept to be benchmarked
- Develop Business cases
- Assess the socio-economical impact
- Disseminate/ Promote /raise public awareness
- Raise engagement in the innovative ecosystem model
- Feed-back loop /return on investment

## CONCLUSION

This paper has looked at some general elements of the RI's of tomorrow. It identifies some new and exciting areas where doors will open to a new potential and an increasing footprint.

But to enable that, the RI's must extend their reach and involve themselves in activities outside of the traditional scientific framework and support the accessibility and distribution of data, as well as innovative ecosystems that both enable and motivate the use of RI results, both in terms of science and engineering.

There are plenty of successful innovative ecosystems in the world where lessons can be learned, and inspiration is fuelled. Look to Boston, Singapore, London, Mumbai, etc. and find some impressive solutions, activities, and results.

For the writers of this paper, the MIT definition of an innovative ecosystem is adapted and here the five stakeholder groups of corporates, governments, entrepreneurs, universities/facilities, and risk/venture capital needs to be represented for a sustainable solution.

Looking at the existing RI I-Ecosystems through these lenses, it becomes more difficult to find an existing reference setup and the conclusion of this paper is, that it is time to approach a full-scale solution to enable the RI potential and impact.

## ACKNOWLEDGEMENTS

The authors thank Fabien Rey, Lars Aprin, Joanna Lewis and Eva Davidsson, for their contributions enhancing collaborative spirits using innovative tools.

# REFERENCES

[1] R. Garoby, "Progress on the ESS Project Construction," in *Proc. 8th Int. Particle Accelerator Conf. (IPAC'17)*, Copenhagen, Denmark, May 2017, pp. 7-12.
   doi:10.18429/JACoW-IPAC2017-MOXBA1

[2] D. N. Argyriou, "Large-science facilities must continue to add value to the scientific community," *IUCrJ*, vol. 6, no. 5, pp. 782–783, Aug. 2019.
   doi:10.1107/s2052252519011709

[3] MAXIV Laboratory, https://www.maxiv.lu.se/

[4] European Organization for Nuclear Research,
   https://home.cern/

[5] ESRF - Research with synchrotron X-rays boosts industrial innovation, https://www.esrf.fr/news/
   general/industrycollaborationsurvey/
   index_html

[6] European Strategy Forum on Research Infrastructures,
   https://www.esfri.eu/

[7] G. Rossi, Roadmap 2018 - European Strategy Forum on Research Infrastructures,
   http://roadmap2018.esfri.eu/

[8] BrightnESS, NEUTRON USERS IN EUROPE: Facility-Based Insights and Scientific Trends,
   https://brightness.esss.se/

[9] European Network of Research Infrastructures & IndusTry for Collaboration, https://enriitc.eu/

[10] Lund institute of advanced neutron and x-ray science,
   https://www.linxs.se

[11] League of European Accelerator-based Photon Sources,
   https://leaps-initiative.eu/

[12] European Research Infrastructure Consortium,
   https://www.ceric-eric.eu/

[13] EIROforum, https://www.eiroforum.org/

[14] League of advanced European Neutron Sources,
   https://www.lens-initiative.org/

[15] Linking Industry to Neutrons and X-rays,
   https://www.linxassociation.com

[16] M. G. J. den Elzen and P. L. Lucas, "The FAIR model: A tool to analyse environmental and costs implications of regimes of future commitments," *Environmental Modeling & Assessment*, vol. 10, no. 2, pp. 115–134, Jun. 2005.
   doi:10.1007/s10666-005-4647-z

[17] Photon and Neutron Open Science Cloud,
   https://www.panosc.eu/

[18] European Open Science Cloud,
   https://eosc-portal.eu/

[19] K. Kaspersen and L. Aaboen, "Feasibility studies at CERN: CERN as a technology provider for startups," *Innovation in Global Entrepreneurship Education*, pp. 196–210, 2021.
   doi:10.4337/9781839104206.00022

[20] ESS Industry and Innovation,
   https://confluence.esss.lu.se/display/INDINN

[21] Non-Invasive Beam Profile Monitor,
   https://brightness.esss.se/sites/default/
   files/Business%20Case_NPM_0.pdf

[22] Data Management & Software Centre,
   https://europeanspallationsource.se/
   data-management-software-centre

[23] S. Petit, "Track It - General presentation," Apr. 2021, EDMS Document n°1884505

[24] The CERN Engineering and Equipment Management System (EDMS),
   https://edms-service.web.cern.ch/
   faq/EDMS/pages/

[25] Final Report Summary - EDUSAFE (Education in advanced VR/AR Safety Systems for Maintenance in Extreme Environments),
   https://cordis.europa.eu/project/id/316919

[26] Nordic Particle Accelerator School,
   https://npaperasmusplus.wordpress.com/

[27] N. Delerue *et al.,* "A Massive Open Online Course on Particle Accelerators," *Journal of Physics: Conference Series*, vol. 1067, p. 092004, Sep. 2018.
   doi:10.1088/1742-6596/1067/9/092004

[28] African School of Fundamental Physics and Applications,
   https://www.africanschoolofphysics.org/

[29] ASP Online courses LS and NS,
   https://www.linxs.se/
   related-events/2020/11/24/
   asp-online-lecture-series-the-african-
   school-of-physics-synchrotron-and-neutron-
   based-diffraction-and-spectroscopic-tech-
   niques

[30] Forum on International Physics/ PHYSICS MATTERS series, https://engage.aps.org/fip/resources/
   activities/physics-matters

[31] SCIENTÍFika, https://www.maxiv.lu.se/
   science/scientifika/